\documentclass[twocolumn,showpacs,preprintnumbers,superscriptaddress,amsmath,amssymb]{revtex4-2}
\usepackage{amssymb}		
\usepackage{amsmath}
\usepackage{graphicx}
\usepackage[normalem]{ulem}
\usepackage{multirow}
\usepackage{appendix}
\usepackage{CJK}
\usepackage[usenames]{color}
\usepackage{bm}
\usepackage{placeins} 
\usepackage[colorlinks,linkcolor=blue,urlcolor=blue,citecolor=blue]{hyperref}
\usepackage{booktabs} 	
\usepackage{leftidx}
\usepackage[utf8]{inputenc}

\usepackage{soul,xcolor}
\setstcolor{red}

\usepackage[all]{hypcap} 
\makeatletter

\newcommand{\Rmnum}[1]{\expandafter\@slowromancap\romannumeral #1@}
\makeatother
\renewcommand{\sout}{\bgroup \color{red} \ULdepth=-.5ex \ULset}

\newcommand\redsout{\bgroup\markoverwith{\textcolor{red}{\rule[0.5ex]{2pt}{0.4pt}}}\ULon}

\usepackage{tikz,xcolor,hyperref}

\definecolor{lime}{HTML}{A6CE39}
\DeclareRobustCommand{\orcidicon}{
	\begin{tikzpicture}
	\draw[lime, fill=lime] (0,0) 
	circle [radius=0.16] 
	node[white] {{\fontfamily{qag}\selectfont \tiny ID}};
	\draw[white, fill=white] (-0.0625,0.095) 
	circle [radius=0.007];
	\end{tikzpicture}
	\hspace{-2mm}
}
\foreach \x in {A, ..., Z}{%
	\expandafter\xdef\csname orcid\x\endcsname{\noexpand\href{https://orcid.org/\csname orcidauthor\x\endcsname}{\noexpand\orcidicon}}
}

\begin{document}
\begin{CJK*}{UTF8}{gbsn}



\title{Production of light nuclei in isobaric Ru + Ru and Zr + Zr collisions at $\sqrt{s_{\mathrm{NN}}}$ =7.7 - 200 GeV from a multiphase transport model}

\author{Fei Li(李飞)}
 \affiliation{Key Laboratory of Nuclear Physics and Ion-beam Application (MOE), Institute of Modern Physics, Fudan University, Shanghai 200433, China}

 \author{Song Zhang(张松)\orcidC{}}
 \email{song$_$zhang@fudan.edu.cn}
 \affiliation{Key Laboratory of Nuclear Physics and Ion-beam Application (MOE), Institute of Modern Physics, Fudan University, Shanghai 200433, China}
  \affiliation{ Shanghai Research Center for Theoretical Nuclear Physics， NSFC and Fudan University, Shanghai 200438, China}

 \author{Kai-Jia Sun(孙开佳)\orcidD{}}
 \email{kjsun@fudan.edu.cn}
 \affiliation{Key Laboratory of Nuclear Physics and Ion-beam Application (MOE), Institute of Modern Physics, Fudan University, Shanghai 200433, China}
  \affiliation{ Shanghai Research Center for Theoretical Nuclear Physics， NSFC and Fudan University, Shanghai 200438, China}

   \author{Yu-Gang Ma(马余刚)\orcidB{}}%
 \email{mayugang@fudan.edu.cn}
 \affiliation{Key Laboratory of Nuclear Physics and Ion-beam Application (MOE), Institute of Modern Physics, Fudan University, Shanghai 200433, China}
 \affiliation{ Shanghai Research Center for Theoretical Nuclear Physics， NSFC and Fudan University, Shanghai 200438, China}

\date{\today}

\begin{abstract}
{The production of light nuclei in isobaric \texorpdfstring{$^{96}_{44}$Ru + $^{96}_{44}$Ru}{96_44Ru + 96_44Ru} and \texorpdfstring{$^{96}_{40}$Zr + $^{96}_{40}$Zr}{96_40Zr + 96_40Zr} collisions, ranging from \texorpdfstring{$\sqrt{s_{NN}}=$7.7}{sNN=7.7} to 200 GeV, are studied using the string melting version of A Multi Phase Transport (AMPT) model in combination with a coalescence approach to light nuclei production. From the calculated yields, transverse momentum (\texorpdfstring{$p_{T}$}{pt}) spectra, and rapidity dependences of light nuclei (\texorpdfstring{$p$, $n$, $d$, $t$, ${}^{3}$He}{p,n,d,t,he}), we find that the Ru+Ru/Zr+Zr ratios for the yields of these particles exceed unity with the inclusion of a quadrupole deformation \texorpdfstring{$\beta_{ 2 }$}{beta2} and octupole deformation \texorpdfstring{$\beta_{ 3 }$}{beta3} as well as the neutron skins. We also find that heavier particles have a larger deviation from unity.
Furthermore, we find that as the collision energy increases, the influence of isospin effects on the production of light nuclei in isobar collisions gradually decreases, while the influence of nuclear structure becomes more significant, particularly evident from the energy dependence of the deuteron ratio, which is unaffected by isospin effects.
}
\end{abstract}
\maketitle


\section{Introduction}
The study of light nuclei production in heavy-ion collisions has received increased interest both experimentally~\cite{PhysRevLett.5.19, adam2019beam, zhang2021light, chen2018antinuclei, ono2019dynamics,Zhang1} and theoretically~\cite{oliinychenko2021overview, braun2019loosely, csernai1986entropy, sun2018light, dover1991covariant,Sun1,Chen1} due to its relevance in probing the critical point of strongly interacting matter~\cite{star2023beam, sun2017probing, shuryak2020baryon, yu2020search, shao2020probing, deng2020light, zhao2020beam, liu2020light,Ko1,He1,Ma1} and indirect dark matter searches~\cite{vagelli2019italian, blum2017cosmic}.

Light nuclei with baryon number $B \leq 4$, such as deuteron ($d$), triton ($t$), and helium-3 ($^3$He), are loosely bound states with binding energies significantly smaller than the temperature of the hot and dense matter created in relativistic heavy-ion collisions. Their production has been analyzed through various models, including the statistical hadronization model~\cite{cleymans1993thermal, andronic2011production, braun2016properties}, the nucleon coalescence model~\cite{sun2018light, butler1961deuterons, sato1981coalescence, PhysRevC.95.044905,shah2016production,Yan1,WangTT,ZhangYX,LiuC,ChengYL,Qiao1}, and dynamical models based on the kinetic theory~\cite{danielewicz1991production}. In the statistical hadronization model, both light nuclei and hadrons are produced at the chemical freeze-out, characterized by a specific temperature and baryon chemical potential. In contrast, the nucleon coalescence model suggests light nuclei formation from kinetically freeze-out nucleons that are close in phase space. Dynamical models treat light nuclei as explicit degrees of freedom, with their dissociation and regeneration controlled by hadronic reactions satisfying the principle of detailed balance~\cite{oh2009deuteron, cho2018kinetic, oliinychenko2019microscopic,WangRui,SunKJ_NC}. Despite the great progress in recent years, the formation mechanism of light nuclei remains under debate.

Isobaric collisions like $^{96}_{44}$Ru + $^{96}_{44}$Ru and $^{96}_{40}$Zr + $^{96}_{40}$Zr~\cite{abdallah2021search,PhysRevC.94.041901,tribedy2020status} collisions aimed at the search of the anomalous phenomenon of chiral magnetic effect (CME) \cite{NRN_Liao,PhysRevC.105.014901,ShenDY,WangCZ}.
They also provide a good opportunity to investigate the impact of initial nuclear structure through its influence on final observables such as $N_{ch}$, $v_2$, $v_3$, etc~\cite{xu2023probing, PhysRevC.107.064909, xi2023vorticity, ZHAO2023137838, PhysRevC.108.L011902, PhysRevLett.131.022301, NIE2023138177, PhysRevC.108.024911}. 
Many studies have shown that the Ru nucleus has quadrupole deformation, while the Zr nucleus has octupole deformation, and they possess different nuclear density~\cite{jia2023scaling,XU2021136453}. 
This difference explains the observed differences in $v_2$ and $v_3$ ratios between the two isobaric $^{96}_{44}$Ru + $^{96}_{44}$Ru and $^{96}_{40}$Zr + $^{96}_{40}$Zr collisions at $\sqrt{s_{NN}}$ = 200 GeV, as recently reported in the results from the STAR experiment~\cite{abdallah2021search}. 
In previous study on the production of light nuclei in isobar collisions, researcher employed $^{96}_{44}$Ru + $^{96}_{44}$Ru and $^{96}_{40}$Zr + $^{96}_{40}$Zr (under the assumption of identical nuclear structures) collisions to investigate the impact of the CME on the production of (anti-)hypernuclei and light (anti-)nuclei, and the results revealed no discernible effect~\cite{she2021predictions}.

In the present study, we aim to explore the effects of initial nuclear structure on the production of light nuclei in isobar collisions by taking into account the different internal structures of $^{96}_{44}$Ru and $^{96}_{40}$Zr nuclei.
Specifically, we investigate light nuclei ($d$, $t$, $^3$He) production at mid-rapidity ($|y| < 0.5$) in isobaric ${}^{96}_{44}$Ru + ${}^{96}_{44}$Ru and ${}^{96}_{40}$Zr + ${}^{96}_{40}$Zr collisions from 7.7 to 200 GeV using a coalescence approach. We explore the impact of nuclear structure and isospin on various final-state observables by employing four different sets of nuclear structure parameters for Ru and Zr. We find that the Ru+Ru/Zr+Zr ratios for yields of these particles exceed unity after accounting for a quadrupole deformation $\beta_{ 2 }$ and octupole deformation $\beta_{ 3 }$ as well as the neutron skin between these two isobars. We also find that the impact of isospin effects on light nuclei production in isobar collisions becomes weak as increasing the collision energy, while the influence of nuclear structure exhibits an opposite trend.

The paper is organized as follows: In Sect. II, the AMPT model with a coalescence approach and four sets of WS parameters for Ru and Zr are briefly introduced. In Sect. III, the analysis of $p_{T}$ spectra, dN/dy, particle ratios for light nuclei in isobar collisions at $\sqrt{s_{NN}}$ = 7.7, 27, 62.4 and 200 GeV are presented. Finally, the summary is given in Sec. IV.

\section{Methods}
\subsection{The AMPT model and nucleon coalescence model} 
To study the production of deuteron ($d$), triton ($t$), and helium-3 ($^3$He) in Ru + Ru and Zr + Zr collisions at $\sqrt{ s_{NN}}$ = 7.7, 27, 62.4 and 200 GeV, we employ the nucleon coalescence model for cluster production with the nucleon phase-space information generated from the string melting version of A Multi-Phase Transport (AMPT) model~\cite{PhysRevC.72.064901,AMPT2021}.
The AMPT model is widely used to study a variety of observables in relativistic heavy-ion collisions. It consists of four parts: the Heavy-Ion Jet INteraction Generator (HIJING) model~\cite{PhysRevD.44.3501, GYULASSY1994307} for generating the initial-state conditions, Zhang's parton cascade (ZPC) model~\cite{ZHANG1998193} for modeling the partonic cascade, the Lund string fragmentation model or a quark coalescence model for hadron formation, and a relativistic transport (ART) model~\cite{PhysRevC.52.2037} for hadronic scatterings and decays. 
 
In this work, we adopt a coalescence approach~\cite{deng2020light, sombun2019deuteron}, in which light nuclei are formed from kinetically freeze-out nucleons that are nearby in phase space. 
In this approach, deuteron ($d$) are formed if their relative coordinate $\Delta r_{p,n}$ and momentum $\Delta p_{p,n}$ between proton ($p$) and neutron ($n$) in the two-nucleon rest frame satisfy $\Delta r_{p,n} \le 3.0$ fm and $\Delta p_{p,n} \le 0.16$ GeV/$c$ simultaneously, where the $\Delta r_{p,n}$ and $\Delta p_{p,n}$ are defined as: 
\begin{equation}
\Delta r_{p,n} = |\mathbf{r}_{p} - \mathbf{r}_{n}|,\quad
\Delta p_{p,n} = |\mathbf{p}_{p} - \mathbf{p}_{n}|,
\end{equation} 
with the $\mathbf{r}_{i}$ and $\mathbf{p}_{i}$ being the coordinate and momentum of nucleon $i$, respectively.
In the two-nucleon rest frame, we synchronize the freeze-out times by allowing the nucleon that freezes out earlier to freely propagate to the later freeze-out time of the other nucleon. 
Similarly, for the triton ($t$) and helium-3 ($^3$He) of the 3-body system, the coalescence parameters are determined by the maximum relative distance $\Delta r_{\max}$ of the nucleon to the coordinates of the center-of-mass and the maximum relative momentum $\Delta p_{\max}$ among the nucleons. The center-of-mass coordinate $\mathbf{R}$ and the maximum relative distance $\Delta r_{\max}$ are given by:
\begin{equation} 
\mathbf{R} = \frac{\mathbf{r}_{1} + \mathbf{r}_{2} +\mathbf{r}_{3}}{3},\quad
\Delta r_{\max} = \max\{|\mathbf{R} - \mathbf{r}_{i}|\}\quad(i = 1, 2, 3),
\end{equation}
and the maximum relative momentum $\Delta p_{\max}$ is given by:
\begin{equation} 
\Delta p_{\max} = \max \{ |\mathbf{p}_i - \mathbf{p}_j| \} \quad (i \neq j; i, j = 1, 2, 3),
\end{equation}
where the $\Delta r_{\max}$ $\le 2.0$ fm, and $\Delta p_{\max}$ $\le 0.16$ GeV/$c$, respectively.

Under the conditions described above, deuterons ($pn$$ \rightarrow $ $d$) are formed by one proton and one neutron, tritons ($pnn$$ \rightarrow $ $t$) are composed of one proton and two neutrons, while helium-3 ($ppn$ $ \rightarrow $ $^3$He) consist of two protons and one neutron. This simple coalescence model is found to describe successfully the experimental data in collisions considered in the present study.

\subsection{Geometric description of \texorpdfstring{$^{96}_{44}$Ru and $^{96}_{40}$Zr}{96-44Ru and 96-40Zr}}

In AMPT, the spatial distribution of nucleons within $^{96}_{44}$Ru and $^{96}_{40}$Zr is characterized by the Woods-Saxon (WS) distribution ~\cite{PhysRevC.94.041901,PRL2021,ShouPLB} function as follows:
\begin{equation}
 \rho ( r , \theta ) = \frac { \rho _ { 0 } } { 1 + \operatorname { exp }\left [ \frac{ r - R _ { 0 } \left(1+ \beta _ { 2 } Y _ { 2 } ^ { 0 } ( \theta ) + \beta _ { 3 } Y _ { 3 } ^ { 0 } ( \theta \right) } {a} \right] },
 \label{eq:nuclei-density}
 \end{equation}
where \textit{r} is radial distance from the center and $\theta$ is polar angle in spherical coordinates. The parameters $R_{0}$ and $a$ represent the "radius" of the nucleus and the surface diffuseness parameter, respectively.   The value of $\rho_{0}$ in Eq. \eqref{eq:nuclei-density} is determined from the nucleon number $A$.  The nuclear shape deformations are characterized by the axial symmetric quadrupole deformation parameter $\beta_2$ and the octupole deformation parameter $\beta_3$, which are the most significant for describing the nucleus's deformation.
 
In the present study, we utilize four sets of Woods-Saxon parameters to characterize the initial nucleon distribution in isobars. These parameters are widely recognized and have been adopted from a variety of recent studies~\cite{RAMAN20011,PRITYCHENKO20161,MOLLER1995185,XU2021136453, jia2023scaling, nijs2021inferring, PhysRevC.94.041901, PhysRevC.98.054907,PhysRevC.101.061901}. The specific values for these parameters are listed in Table~\ref{tbl1}. This approach allows us to systematically investigate the impact of nuclear structure variations on the production of light nuclei in isobaric collisions.
  
 \begin{table}[!t] 
 \centering
 \caption{The Woods-Saxon parameters used in the AMPT model.}
 \label{tbl1}
 \fontsize{9}{20}
 \renewcommand\arraystretch{1.2}
 \begin{tabular} {|c|c|c|c|c|c|c|c|c|}
  \hline
 &\multicolumn{4}{c|}{$^{96}_{44}$Ru} &\multicolumn{4}{c|}{$^{96}_{40}$Zr}\cr\cline{2-9}
 Nucleus	  & $R_{0}$~(fm) & $a$~(fm)  & $\beta_{2}$  & $\beta_{3}$ 
 		& $R_{0}$~(fm) & $a$~(fm) & $\beta_{2}$  & $\beta_{3}$  \cr
 		\hline
  Case-1
 		& 5.096& 0.540 & 0& 0 
	  & 5.096& 0.540 & 0& 0 \cr\cline{2-9}\hline 		
  Case-2
 		& 5.13& 0.46& 0.13& 0 
 		& 5.06& 0.46 & 0.06& 0 \cr\cline{2-9}\hline
  Case-3
 		& 5.067& 0.5& 0& 0
 		& 4.965 & 0.556 & 0& 0 \cr\cline{2-9}\hline
  Case-4
 		& 5.065& 0.485& 0.154& 0
 		& 4.961 & 0.544 & 0.062& 0.202 \cr\cline{2-9}\hline
 \end{tabular}
 \end{table}

The parameters for the initial nucleon distributions are organized into four cases: Case-1 assumes identical values for $a$ and $R_{0}$ for both Ru and Zr, with no deformation present in either nucleus; Case-2 differentiates the two nuclei by their quadrupole deformation, with Ru ($\beta^{Ru}_{2}$ = 0.13) exhibiting a larger deformation than Zr ($\beta^{Zr}_{2}$ = 0.06); Case-3, as referenced from \cite{XU2021136453,PhysRevLett.121.022301}, posits both nuclei as spherical ($\beta_{2}$ = 0). Here, variations in $R_{0}$ and $a$ suggest that Ru is overall smaller than Zr, largely due to Zr having a more substantial neutron skin;
Case-4 provides a comprehensive description of the nuclear structure, encompassing both the deformation effect ($\beta_2$ and $\beta_3$)~\cite{jia2023scaling,nijs2021inferring} and the neutron skin effect for $^{96}_{44}$Ru and $^{96}_{40}$Zr, which leads to the best description of $v_{2,Ru}$/$v_{2,Zr}$ in isobar collisions at $\sqrt{ s_{NN}}$ = 200 GeV~\cite{PhysRevC.106.014906}.
It should be noted that we use the total nucleon density (the sum of the proton and neutron) to represent the neutron skin effect caused by different density distributions of proton and neutron. The feasibility of this approach is supported by previous research\cite{XU2021136453}, which has shown that the Ru+Ru/Zr+Zr ratio remains consistent when comparing calculations using distinct proton and neutron densities in energy density functional theory (DFT) calculations with those using total nucleon density in WS parameters.

 \FloatBarrier
\begin{figure*}[!t]
 \centering
 \includegraphics[scale=.28]{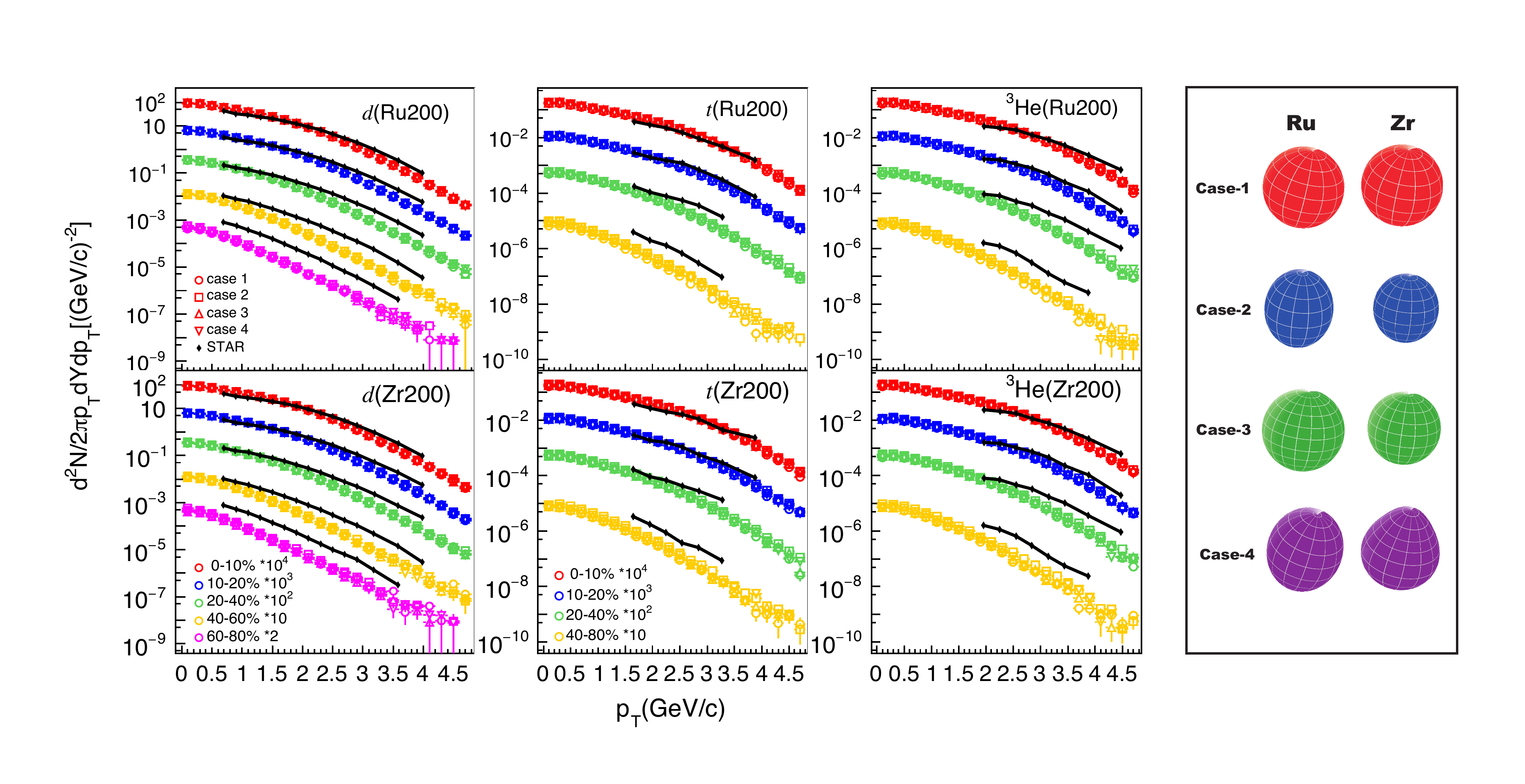} 
 \caption{Transverse momentum spectra for $d$ ($|y|<$0.35), $t$ ($|y|<$0.5), and $^3$He ($|y|<$0.5) in 0-10\%, 10-20\%, 20-40\%, 40-60\%, and 60-80\% (40-80\% for $t$ and $^3$He) isobaric collisions at $\sqrt{ s_{NN}}$ = 200 GeV with four different sets of Woods-Saxon parameters. The STAR data (black solid lines) for isobar collisions at $\sqrt{ s_{NN}}$ = 200 GeV~\cite{liu2022measurement} are also shown for comparisons.}
 \label{FIG:fig1}
\end{figure*}

\section{RESULTS AND DISCUSSION}
The results from the AMPT model with a coalescence approach for four sets of WS parameters in Ru + Ru and Zr + Zr collisions at $\sqrt{ s_{NN}}$ = 7.7, 27, 62.4 and 200 GeV are presented, as well as for the STAR data on isobar collisions at $\sqrt{ s_{NN}}$ = 200 GeV for comparisons. We will show the predictions for the transverse momentum $p_{T}$ spectra, the particle yield $dN/dy$, and the particle ratios of light nuclei as a function of centrality at mid-rapidity in isobar collisions. We will also discuss the energy dependence of the effects of isospin and nuclear structure on the ratios of these observations between the two collision systems. The numbers of events we simulated for all cases for both $^{96}_{44}$Ru and $^{96}_{40}$Zr are, respectively, 5M (7.7 GeV), 4M (27 GeV), 3.5M (62.4 GeV), and 5M (200 GeV), where M denotes $\times 10^6$.

\begin{figure*}
\centering
\includegraphics[scale=.9]{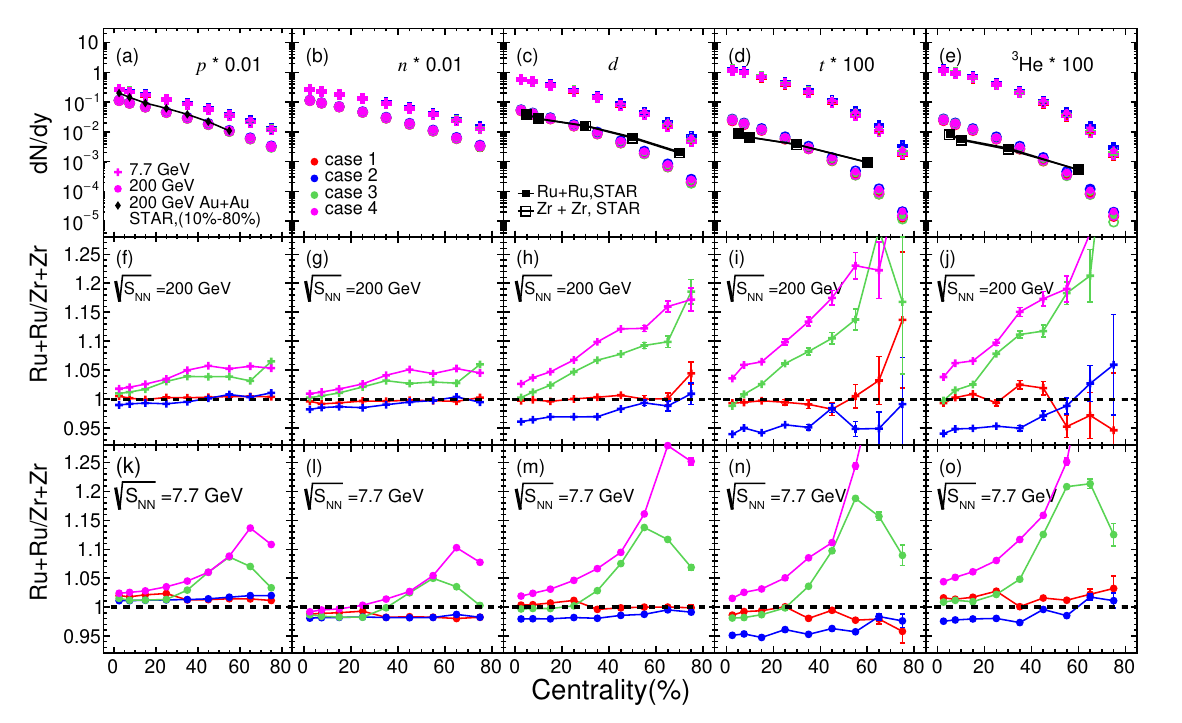} 
\caption{
  The rapidity density ($dN/dy$) of $p$, $n$, $d$, $t$, and $^3$He at mid-rapidity ($|y|<$0.5) versus the centrality in Ru + Ru and Zr + Zr collisions at $\sqrt{s_{NN}}$ = 7.7 and 200 GeV for four cases of WS parameters. dN/dy of $p$, $n$ and $t$, $^3$He are scaled with factors of 0.01 and 100 for better representation, respectively.  The STAR data for $d$, $t$, and $^3$He in isobar collisions~\cite{liu2022measurement}, as well as $p$ from Au + Au collisions (10-80\%) \cite{PhysRevC.79.034909} at $\sqrt{s_{NN}} = 200$ GeV, are also shown for comparisons.
  Ru is represented by solid symbols and Zr by open symbols. The Ru+Ru/Zr+Zr ratio of dN/dy is shown in the lower panels for $\sqrt{s_{NN}}$ = 7.7 and 200 GeV, respectively.}
  
\label{FIG:fig2}
\end{figure*}

\subsection{The \texorpdfstring{$p_{T}$}{pT} spectra of light nuclei}

Shown in figure \ref{FIG:fig1} are the transverse momentum $p_{T}$ spectra of deuterons ($|y|<$0.35), tritons ($|y|<$0.5), and helium-3 ($|y|<$0.5) in Ru + Ru and Zr + Zr collisions. The spectra are shown across various centrality bins: 0-10\%, 10-20\%, 20-40\%, 40-60\%, and 60-80\% (40-80\% for $t$ and $^3$He) using four different WS parameter sets. For comparison, the STAR data on these light nuclei in isobar collisions at $\sqrt{s_{NN}}=200$ GeV is also included.

The $p_{T}$ spectra of deuterons ($d$), tritons ($t$), and helium-3 ($^3$He) with four different sets of WS parameters agree well with STAR data in the central collision region(0-10\% and 10-20\%). However, in the more peripheral collisions (40-60\% and 60-80\%), the obtained yields underestimate the STAR data. 
The underestimation of light nuclei production in peripheral collisions results from the imperfection of the hadron production mechanism in the AMPT model in peripheral collisions or small system collisions, which needs further improvement but is beyond the scope of this work. Fortunately, this issue does not affect the following defined ratios significantly.

\begin{figure*}
	\centering
	\includegraphics[scale=.9]{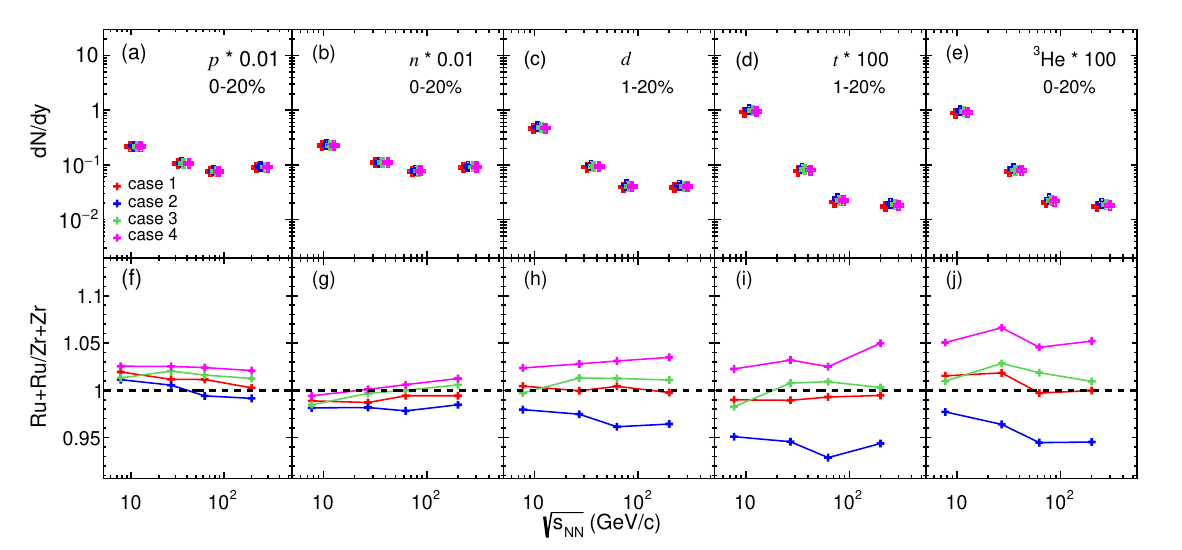} 
	\caption{
		The rapidity density ($dN/dy$) of $p$, $n$, $d$, $t$, and $^3$He at mid-rapidity ($|y|<$0.5) in the central isobar collisions as a function of energy for four cases of WS parameters. $dN/dy$ of $p$, $n$, $t$, and $^3$He are scaled with factors of 0.01 and 100 for better representation, respectively. Ru is represented by solid symbols and Zr by open symbols. The Ru+Ru/Zr+Zr ratio of $dN/dy$ is shown in the lower panels. 
	}
	\label{FIG:fig3}
\end{figure*}

\subsection{The yields of light nuclei}

Figure \ref{FIG:fig2} depicts the yields ($dN/dy$) of $p$, $n$, $d$, $t$, and $^3$He at mid-rapidity ($|y|<$0.5) as a function of centrality in Ru + Ru and Zr + Zr collisions at $\sqrt{ s_{NN}}$ = 7.7 and 200 GeV, utilizing four sets of WS parameters. 
For subsequent analysis, the rapidity of deuterons is constrained to $|y|<$0.5.
Additionally, the STAR data on the rapidity density ($dN/dy$) for $d$, $t$, and $^3$He in isobar collisions~\cite{liu2022measurement}, as well as protons from Au + Au collisions~\cite{PhysRevC.79.034909}, selected from a similar multiplicity collision region (10-80\%) as isobar collisions, at $\sqrt{s_{NN}} = 200$ GeV, are presented for comparison.
With the four sets of WS parameters, both the $dN/dy$ of each particle are consistent between Ru + Ru and Zr + Zr collisions at $\sqrt{ s_{NN}}$ = 7.7 and 200 GeV. The $dN/dy$ of four cases all exhibit a decreasing trend from central to peripheral collisions at both collision energies. 

The yields ($dN/dy$) of light nuclei are shown in the upper panels and the Ru+Ru/Zr+Zr ratio of $dN/dy$ for $p$, $n$, $d$, $t$, and $^3$He at $\sqrt{ s_{NN}}$ = 200 and 7.7 GeV in the middle panels and the lower panels respectively.
For the Ru+Ru/Zr+Zr ratios of $p$ and $n$ (as illustrated in Fig. \ref{FIG:fig2}~(f) and Fig. \ref{FIG:fig2}~(g)), their values remain near unity across all centralities at 200 GeV in case 1, where the $^{96}_{44}$Ru and $^{96}_{40}$Zr nuclei are the same nuclear size and without deformation. 
In case 2, which incorporates larger quadrupole deformation of $^{96}_{44}$Ru compared to $^{96}_{40}$Zr, the Ru+Ru/Zr+Zr ratios of $p$ and $n$ slightly dip below unity within central collision regions at 200 GeV. 
Conversely, in Cases 3 and 4, the Ru+Ru/Zr+Zr ratios of $p$ and $n$ ratios almost linearly increase with centrality. Specifically, the ratios in case 4, which considers both quadrupole and octupole deformation as well as the neutron skin effect, exceed those in case 3 which accounts only for the neutron skin effect.
For $d$, $t$, and $^3$He, depicted in Fig.\ref{FIG:fig2}~(h), (i), and (j), their Ru+Ru/Zr+Zr ratios exhibit a similar trend to those of $p$ and $n$ but with more pronounced deviations from unity in cases 2, 3, and 4.

The reason for a larger particle yield in $^{96}_{44}$Ru + $^{96}_{44}$Ru collisions compared to $^{96}_{40}$Zr + $^{96}_{40}$Zr collisions in case 3 and case 4, both of which involve the neutron skin effect, is that $^{96}_{44}$Ru nuclei has a tighter nuclear geometry than $^{96}_{40}$Zr nuclei in the non-peripheral collision region, as the size of $^{96}_{44}$Ru is overall smaller than that of $^{96}_{40}$Zr. This leads to more participating nucleons in $^{96}_{44}$Ru + $^{96}_{44}$Ru collisions than $^{96}_{40}$Zr + $^{96}_{40}$Zr collisions, and as we know the yield of the final particle is related to participating nucleon number $N_{part}$~\cite{jia2023scaling,XU2021136453}. 
Comparing Case 2 (which includes deformation effects) with Case 1 (which has no nuclear structure effect), and Case 4 (which includes both deformation and neutron skin effects) with Case 3 (which includes only the neutron skin effect), it is clear that the trend in the ratios reflects the yield differences in the isobar system caused by pure deformation effects. In this system, nuclei with more deformation have lower yields, indicating that the greater the deformation of the nucleus, the smaller the initial nucleon number. This is consistent with the results in~\cite{abdallah2021search,jia2023scaling,XU2021136453}

At a lower energy of $\sqrt{ s_{NN}}$ = 7.7 GeV, the proton ($p$) ratio mostly increases for all cases relative to the results at 200 GeV, while the neutron ($n$) ratio mostly decreases relative to the result at 200 GeV, as shown in Fig. 2(k) and (l). This is different from the trend observed at 200 GeV, where the $p$ and $n$ ratios were almost identical.
Compared to the initial $^{96}_{44}$Ru + $^{96}_{44}$Ru collision system, the initial $^{96}_{40}$Zr + $^{96}_{40}$Zr collision system had a higher neutron-to-proton ratio ($n/p$).
The difference of Ru+Ru/Zr+Zr ratio between the $p$ and $n$ at different energies indicates the isospin effect on the yield ratio of $p$ and $n$ at lower collision energy have a more significant impact, and this influence is attenuated at higher energy.

\subsection{The yields energy dependence of light nuclei}
Figure~\ref{FIG:fig3} presents the yields energy dependence of $p$, $n$, $d$, $t$, and $^3$He in central collisions (0-20\%). 
With increasing energy, a trend is observed where the ratio for protons ($p$) in case 1 decreases towards unity, while the ratio for neutrons ($n$) increases towards unity. This suggests an amplification of the isospin effect in isobar collisions at lower energies.

Additionally, as shown in Fig.~\ref{FIG:fig3}(h), the ratio for deuterons ($d$, consisting of $p$-$n$ pairs) in case 1 remains constant at unity across different energies, suggesting that the production mechanism of $p$-$n$ paired for deuteron is unaffected by the isospin differences between $^{96}_{44}$Ru + $^{96}_{44}$Ru and $^{96}_{40}$Zr + $^{96}_{40}$Zr collisions. 
Conversely, as depicted in Fig.\ref{FIG:fig3}(i), the triton ($t$, consisting of $pnn$) ratio in case 1 exhibits a trend similar to that of neutrons, while Fig.\ref{FIG:fig3}(j) demonstrates that the ratio for helium-3 ($^3$He, consisting of $ppn$) in case 1 follows the trend of protons, implying their Ru+Ru/Zr+Zr ratio are influenced by the excess nucleon relative to the $p$-$n$ pair.

Considering that the Ru+Ru/Zr+Zr ratio of deuteron is not affected by isospin effects in case 1, where $^{96}_{44}$Ru and $^{96}_{40}$Zr nuclei have the same nuclear structure, the additional cases depicted in Fig. \ref{FIG:fig3}(h) offer the potential to understand the energy dependence of the nuclear structure on light nuclei production.
In cases 2 and 4, the effect of nuclear structure on yield ratio increases with energy. Case 3 displays nonmonotonicity; this is because the neutron skin effect's influence on yield ratios is predominantly observed in mid-central collisions~\cite{PhysRevLett.128.022301, jia2023scaling, PhysRevC.106.014906}. Therefore, the results for case 3 in the central collision region may not accurately reflect the energy dependence of the neutron skin effect.

Furthermore, in Fig. \ref{FIG:fig3}(f), the energy dependence of the ratios for protons ($p$) in case 4 remains almost constant. This is attributed to the interplay between nuclear structure and isospin effects at varying energies, with their energy dependencies exhibiting contrasting behaviors.

 \begin{figure*}
 	\centering
 	\includegraphics[scale=.85]{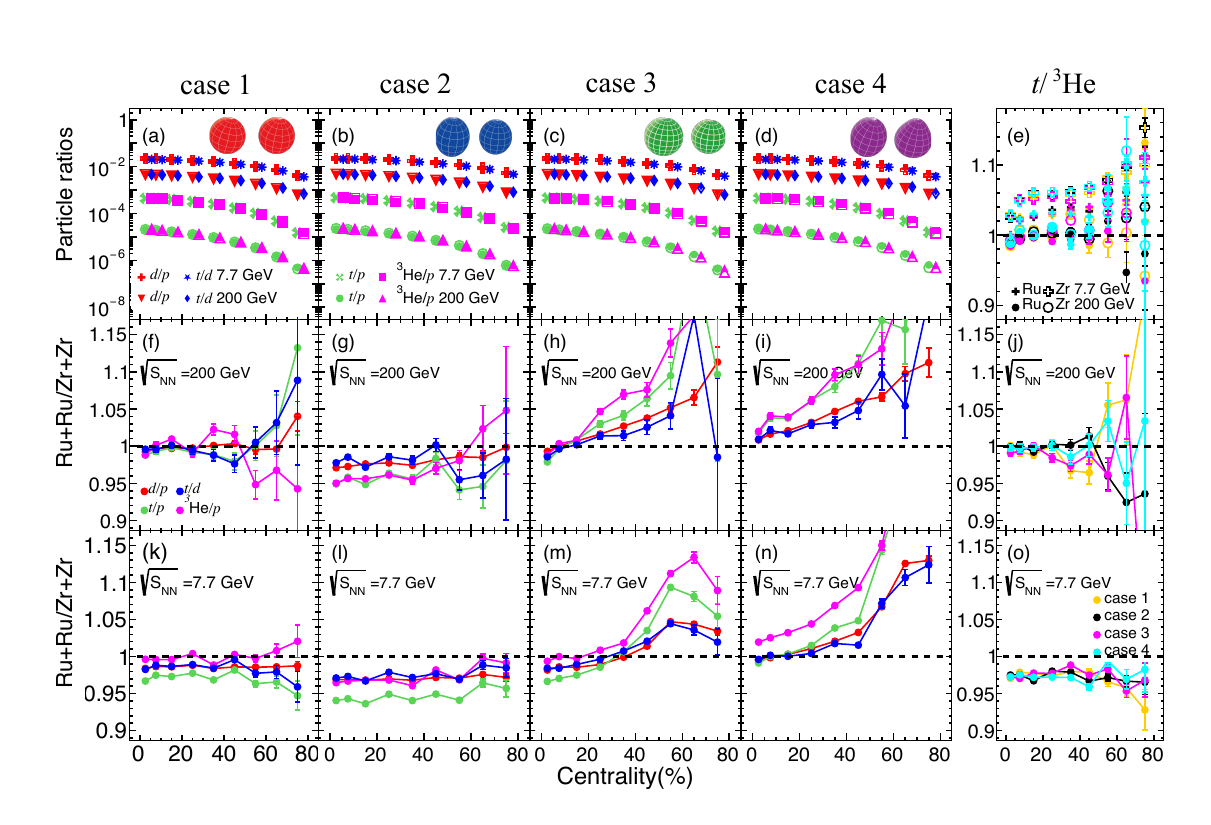} 
 	\caption{
 		The particle ratios of $d/p$, $t/p$, $^3$He/$p$, $t/d$ and $t$/$^3$He at mid-rapidity ($|y|<$0.5) versus the centrality in Ru + Ru and Zr + Zr collisions at $\sqrt{s_{NN}}$ = 7.7 and 200 GeV for four cases of WS parameters. Ru is represented by solid symbols and Zr by open symbols. The Ru+Ru/Zr+Zr ratio of particle ratios is shown in the lower panels for $\sqrt{s_{NN}}$ = 7.7 and 200 GeV, respectively. 		
 	}
 	\label{FIG:fig4}
 \end{figure*}
 
\subsection{The ratios of light nuclei}
Figure~\ref{FIG:fig4} depicts the particle ratios $d/p$, $t/p$, $^3$He/$p$, $t/d$ and $t$/$^3$He at mid-rapidity ($|y|<$0.5) verse the centrality for $^{96}_{44}$Ru + $^{96}_{44}$Ru and $^{96}_{40}$Zr + $^{96}_{40}$Zr collisions at $\sqrt{ s_{NN}}$ = 7.7 GeV and 200 GeV, using four sets of WS parameters. The particle ratios $d/p$, $t/p$, $^3$He/$p$, and $t/d$ consistently decrease from central to peripheral collisions for all cases at $\sqrt{ s_{NN}}$ = 7.7 GeV and 200 GeV. 

Notably, for all cases, the particle ratio $d/p$ ($pn/p$) closely aligns with $t/d$ ($pnn/pn$), and similarly, $t/p$ ($pnn/p$) is close to $^3$He/$p$ ($ppn$/p).
This can be attributed to the fact that they correspond to the same nucleon component multiples. Specifically, the ratios of both $d/p$ and $t/d$ correspond to the nucleon component $p$, while $t/p$ and $^3$He/$p$ are associated with $nn$ and $pn$ respectively. 

\begin{figure*}
 	\centering
 	\includegraphics[scale=.85]{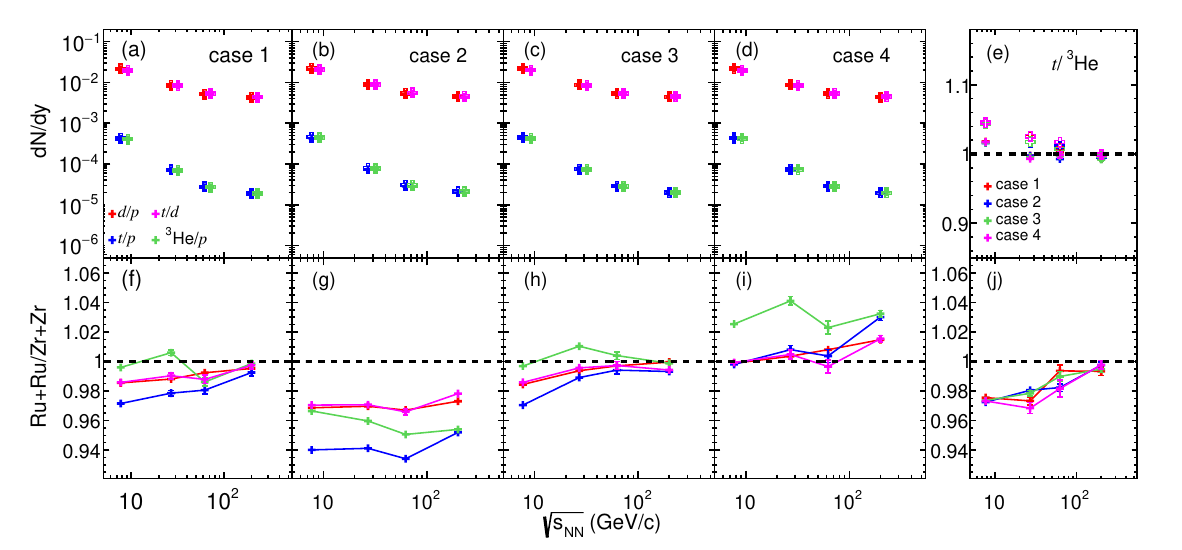} 
 	\caption{
 The particle ratios of $d/p$, $t/p$, $^3$He/$p$, $t/d$, and $t$/$^3$He at mid-rapidity ($|y|<$0.5) in the central isobar collisions as a function of energy for four cases of WS parameters. Ru is represented by solid symbols and Zr by open symbols. The Ru+Ru/Zr+Zr ratio of particle ratios is shown in the lower panels. 	}
 	\label{FIG:fig5}
 \end{figure*}

The lower panels of Fig. \ref{FIG:fig4} depict the ratio between Ru+Ru and Zr+Zr collisions of the above particle ratios at $\sqrt{ s_{NN}}$ = 200 GeV and 7.7 GeV for all cases. 
In case 1, where there are no nuclear structure differences between ${}^{96}_{44}$Ru and ${}^{96}_{40}$Zr, all Ru+Ru/Zr+Zr ratios of particle ratios converge to unity in central collisions at $\sqrt{s_{NN}}$ = 200 GeV. In contrast, variations in these ratios are evident (except for $d/p$ and $t/d$ being the same) across all centralities at $\sqrt{s_{NN}} = 7.7$ GeV. 
The difference between $\sqrt{ s_{NN}}$ = 200 and 7.7 GeV can be attributed to a stronger isospin effect at the lower energy, as we stated before. 
The Ru+Ru/Zr+Zr ratios of $d/p$ and $t/d$ remain the same at the lower energy because they have the same nucleon component $p$, while $t/p$ and $^3$He/$p$ separate at lower energy for they possess different nucleon components.
The above results are also consistent with other cases involving nuclear structure, where the ratios of $d/p$ and $t/d$ consistently remain the same. Conversely, the ratios of $t/p$ and $^3$He/$p$ exhibit a proximity at $\sqrt{ s_{NN}}$ = 200 GeV, while displaying a difference at lower energy of 7.7 GeV. This can be seen more clearly in the results of the energy dependence of yields ratio in Fig. \ref{FIG:fig5}.

In addition, Figure.~\ref{FIG:fig4}(e) shows that the yield ratio $t$/${}^{3}$He is close to unity in central collisions and increases to 1.1 in peripheral collisions. However, we note that this ratio is much larger (about 1.8) in the preliminary result from STAR~\cite{liu2022measurement}, although the uncertainty is significantly large.
In Fig.~\ref{FIG:fig4}(j)(o), the Ru+Ru/Zr+Zr ratios of $t$/$^3$He for four cases of WS parameters at $\sqrt{ s_{NN}}$ =200 GeV and 7.7 GeV are close to each other because of the cancellation of nuclear structure with the same nucleon component for triton and $^3$He. It is also shown in Fig. \ref{FIG:fig5}(j) that the equivalent Ru+Ru/Zr+Zr ratio of $t$/$^3$He converges to unity as the energy increases for all cases. 
This further corroborates that the impact of the initial isospin differences on the isobar collision system weakens with increasing energy.

\section{Conclusion and Outlook}
We investigate $p_{T}$ spectra, $dN/dy$, and particle ratios for $p$, $n$, $d$, $t$ and $^3$He at mid-rapidity ($|y|<$0.5) in ${}^{96}_{44}$Ru + ${}^{96}_{44}$Ru and ${}^{96}_{40}$Zr + ${}^{96}_{40}$Zr collisions at $\sqrt{s_{NN}}$ =7.7, 27, 62.4 and 200 GeV using the AMPT model combined with a final-state coalescence model for light nuclei production.
Using four sets of WS parameters for ${}^{96}_{44}$Ru and ${}^{96}_{44}$Zr, the effect of nuclear structure and isospin on the production of light nuclei in isobaric collisions and the energy dependence of this effect are investigated.

We find that there is little difference in $dN/dy$ and particle ratios for each particle between collisions of ${}^{96}_{44}$Ru + ${}^{96}_{44}$Ru and ${}^{96}_{40}$Zr + ${}^{96}_{40}$Zr at 200 GeV when assuming identical nuclear structures for ${}^{96}_{44}$Ru and ${}^{96}_{40}$Zr. A difference in $dN/dy$ and particle ratios is observed between isobar collisions when these two nuclei possess distinct nuclear structures. A maximum deviation from unity in the Ru+Ru/Zr+Zr ratio of dN/dy and particle ratios is observed for each particle at 200 GeV when considering the nuclear structure with quadrupole deformation $\beta_{2}$, octupole deformation $\beta_{3}$, and neutron skin for ${}^{96}_{44}$Ru and ${}^{96}_{40}$Zr nuclei.
We also found that the heavier particles exhibit greater deviations from unity in the $dN/dy$ and particle ratios between isobar collisions, which is consistent with the results of Ref.~\cite{PhysRevC.108.024911}.

In addition, we find that the impact of initial isospin on light nuclei production in isobar collisions weakens with increasing collision energy, while the effect of nuclear structure exhibits an opposite trend.
This can be seen in the behavior of deuteron in cases 1, 2, and 4 as shown in Fig.~\ref{FIG:fig3}. 
Furthermore, we find that the value of $t$/${}^{3}$He in peripheral collisions is about 1.1 which is, however, much lower than the preliminary result of about 1.8 reported by STAR~\cite{liu2022measurement} although with significant uncertainty.
Our results can be tested in upcoming data from the STAR Collaboration and may provide a deeper understanding of the formation mechanism of light nuclei as well as the effects of internal structures of colliding nuclei.

\begin{acknowledgments}
We thank Xiao-Feng Luo and Hui Liu for fruitful discussions. This work was supported in part by the National Natural Science Foundation of China under contract Nos. 12275054, 11890710, 11890714, 11925502, 12147101, 12061141008, 11875066, and 12375121, National Key R\&D Program of China under Grant No. 2022YFA1602303 and 2018YFE0104600, the Strategic Priority Research Program of CAS under Grant No. XDB34000000, the Guangdong Major Project of Basic and Applied Basic Research No. 2020B0301030008, Shanghai Special Project for Basic Research No. 22TQ006, and the STCSM under Grant No. 23590780100.
\end{acknowledgments}

\nocite{*}

\end{CJK*}

\bibliography{EffectNuclearDeformationLightNuclei}

\end{document}